\documentclass[aps,prd,superscriptaddress,showpacs,linenumbers]{revtex4}
\usepackage{graphicx}
\usepackage{epstopdf}
\usepackage{amsmath}
\usepackage{amsfonts}
\usepackage{amssymb}
\usepackage{latexsym}
\usepackage{lineno}

\begin{document}
\title{Remarks on inverse electrodynamics}
\author{Patricio Gaete} \email{patricio.gaete@usm.cl} 
\affiliation{Departamento de F\'{i}sica and Centro Cient\'{i}fico-Tecnol\'ogico de Valpara\'{i}so-CCTVal,
Universidad T\'{e}cnica Federico Santa Mar\'{i}a, Valpara\'{i}so, Chile}
\author{Jos\'{e} A. Helay\"{e}l-Neto}\email{helayel@cbpf.br}
\affiliation{Centro Brasileiro de Pesquisas F\'{i}sicas (CBPF), Rio de Janeiro, RJ, Brasil} 
\date{\today}

\begin{abstract}
We study physical aspects for a new nonlinear electrodynamics (inverse electrodynamics). 
It is shown that this new electrodynamics displays the vacuum
birefringence phenomenon in the presence of external magnetic field, hence we compute the bending of light. Afterwards we compute the lowest-order modification to the interaction energy within the framework of the gauge-invariant but path-dependent
variables formalism. 
Our calculations show that the interaction energy contains a long-range (${1 \mathord{\left/
 {\vphantom {1 {{r^5}}}} \right.\kern-\nulldelimiterspace} {{r^5}}}$-type)
correction to the Coulomb potential. 
\end{abstract}
\maketitle

\section{Introduction}

Quantum vacuum nonlinearities have a long history originating from the pioneering work by Heisenberg and Euler \cite{H_E}, who obtained an effective nonlinear electromagnetic theory in vacuum arising from the interaction of photons with virtual electron-positron pairs. Subsequently, Schwinger reconfirmed this amazing prediction of light-by-light scattering from Quantum Electrodynamics (QED) \cite{Schwinger}.

In this context it is particularly important to recall that one of the remarkable physical effects of the Heisenberg and Euler result has been vacuum birefringence. More precisely, the quantum vacuum when is stressed by external electromagnetic fields behaves as if it were a birefringent material medium, which has indeed been highlighted from different perspectives \cite{Adler,Costantini,Ruffini,Dunne,Battesti_Rizzo,Sarazin}. However, this optical phenomenon has not yet been confirmed. 

Nevertheless, this remarkable quantum characteristic of light has triggered a growing interest on the experimental side \cite{Bamber,Burke,nphoton,Tommasini1,Tommasini2}. For example, the PVLAS collaboration \cite{Ejilli}. Recently, the ATLAS collaboration has reported on the direct detection of the light-by-light scattering in LHC Pb-Pb collisions \cite{Atlas,Enterria}. The advent of laser facilities has given rise to various proposals to probe quantum vacuum nonlinearities \cite{Battestti2018,Ataman}. More recently, by exploiting the change in the index of refraction due to nonlinear electrodynamics a new experiment (DeLLight project) has been suggested \cite{Robertson2}.

Let us also mention here that different nonlinear electrodynamics of the vacuum may have meaningful contributions to photon-photon scattering such as Born-Infeld  \cite{BI} and Lee-Wick \cite{Lee-Wick1, Lee-Wick2} theories. Recalling that these electrodynamics were introduced in order to avoid the divergences inherent in the Maxwell theory at short distances. In addition, we note that nonlinear electrodynamics have also attracted important attention because they arise naturally in string theories. As is well known, the low energy dynamics of D-branes is described by a Born-Infeld type action \cite{Tseytlin,Gibbons}. Mention should be made, at this point, to a novel electrodynamics (inverse electrodynamics) which has been the object of investigation in the context of Reisner-Nostr\"om black hole solutions \cite{Cembranos}.

On the other hand, previously \cite{Nonlinear,Logarithmic,Nonlinear2,Nonlinear3}, we have examined the physical effects presented by different models of $(3+1)$-D nonlinear electrodynamics in vacuum. This has also helped us to gain insights over the peculiarities of quantum vacuum nonlinearities in different contexts. For example, the Generalized Born-Infeld, and Logarithmic Electrodynamics the field energy of a point-like charge is finite, which also exhibit the vacuum birefringence phenomenon. Additionally, we have studied the lowest-order modifications of the static potential within the framework of the gauge-invariant but path-dependent variables formalism, which is an alternative to the Wilson loop approach. 

From the preceding considerations and given the ongoing experiments related to photon-photon interaction physics, it is useful to further examine the phenomenological consequences presented by a new nonlinear electrodynamics.
 Seem from such a perspective, the present work supplement our previous studies. We also hope that the model discussed here can be helpful in black holes physics. Specifically, we will be concerned with birefringence, bending of light, as well as the computation of the static potential along the lines of \cite{Nonlinear,Logarithmic,Nonlinear2,Nonlinear3}. 

Our work is organized according to the following outline: in Sect. $2$, we address general aspects of this new electrodynamics, show that it yields birefringence, compute the bending angle and calculate the interaction energy for a pair of static probe charges. In Sec. $3$, we compare our previous results with a related model \cite{Cembranos}. Finally, in Sec. $4$, we make final remarks.

In our conventions the signature of the metric is ($+1,-1,-1,-1$). 

\section{The model under consideration}

\subsection{General aspects}

We commence our considerations with a brief description of the model under consideration. The model is characterized by the following Lagrangian density:   
\begin{equation}
{\cal L} =  - {\cal F} \left\{ {1 + \lambda {{\left( {\frac{{{\Lambda ^4}}}{{2{\cal F}}}} \right)}^{2\delta }}} \right\}, \label{INVE05}
\end{equation}
where ${\cal F} = \frac{1}{4}F_{\mu \nu } F^{\mu \nu }$. The constant $\Lambda$ has $mass$ dimension in natural units, whereas $\lambda$ is dimensionless, and $\delta = 1, 2, 3, ...$. Evidently, when $\lambda \to 0$ we recover the Maxwell regime. It is to be specially noted that the mass scale, $M$, fixed by the $\Lambda$-parameter, characterizes a regime where the nonlinearity of the electromagnetic field becomes significant.

The dependence on the field strength in s non-polynomial form
as given in the Lagrangian density of eq.(\ref{INVE05}) appears in connection
with the phenomenon of magnetic catalysis that takes place in the
framework of the Nambu-Jona-Lasinio model. In considering fermions
in presence of magnetic fields in $(1+3)$-dimensional space-time, a chiral condensate is quantum-mechanically generated which, in turn, contributes a term of the form of the one in eq.(\ref{INVE05}) to the effective action \cite{Grasso,Sahu,Miransky}.

From the above Lagrangian density the corresponding equations of motion read
\begin{equation}
\nabla  \cdot {\bf D} = 0, \  \  \
\frac{{\partial {\bf D}}}{{\partial t}} - \nabla  \times {\bf H} = 0, \label{INVE10a}
\end{equation}
\begin{equation}
\nabla  \cdot {\bf B} = 0, \  \  \
\frac{{\partial {\bf B}}}{{\partial t}} + \nabla  \times {\bf E} = 0, \label{INVE10b}
\end{equation}
where the ${\bf D}$ and ${\bf H}$ fields are given by
\begin{equation}
{\bf D} = \left[ {1 + \lambda \left( {1 - 2\delta } \right){{\left( {\frac{{{\Lambda ^4}}}{{\left( {{{\bf B}^2} - {{\bf E}^2}} \right)}}} \right)}^{2\delta }}} \right]{\bf E}, \label{INVE15}
\end{equation}
and
\begin{equation}
{\bf H} = \left[ {1 + \lambda \left( {1 - 2\delta } \right){{\left( {\frac{{{\Lambda ^4}}}{{\left( {{{\bf B}^2} - {{\bf E}^2}} \right)}}} \right)}^{2\delta }}} \right]{\bf B}.  \label{INVE20}
\end{equation}

We first observe that (from Gauss law) for an external point-like charge, q, at the origin, the ${\bf D}$-field lies along the radial direction and is given by ${\bf D} =\frac{q}{{4\pi r^2 }}\hat r$. Making use of this result, it follows that the electrostatic field to leading order in $\lambda$ can be written as
\begin{equation}
|{\bf E}| = \frac{q}{{4\sqrt 2 \pi {r^2}}}\sqrt {1 + \sqrt {1 + \frac{{2048{\pi ^4}\lambda {\Lambda ^8}{r^8}}}{{{q^4}}}} }. \label{INVE25}
\end{equation}

Second, from the foregoing  equations of motion it follows that the electromagnetic vacuum behaves like a polarizable medium, as a consequence optical properties can be explored. In fact, in previous work \cite{Nonlinear,Logarithmic,Nonlinear2,Nonlinear3}, we have considered the phenomenon of optical birefringence, which refers to the property that polarized light in a particular direction (optical axis) travels at a different velocity from that of light polarized in a direction perpendicular to this axis. Following analysis similar to that in \cite{Nonlinear,Logarithmic,Nonlinear2,Nonlinear3}, we now consider a weak electromagnetic wave $({\bf E_p}, {\bf B_p})$ propagating in the presence of a strong constant external field $({\bf E_0}, {\bf B_0})$. In passing we note that, for computational simplicity, we will only consider the case of a purely external magnetic field, namely, ${\bf E_0}=0$. Accordingly, one easily finds  
\begin{equation}
{\bf D} = \xi \, {\bf E}_{p}, \label{INVE30}
\end{equation}
and
\begin{equation}
{\bf H} = \xi \left[ {{{\bf B}_p} + \frac{\chi }{\xi }\frac{{\left( {{{\bf B}_p} \cdot {\bf B_0}} \right)}}{{{\bf B}_0^2}}{{\bf B_0}}} \right], \label{INVE35} 
\end{equation}
with
\begin{equation}
\xi  = 1 + \lambda \left( {1 - 2\delta } \right){\left( {\frac{{{\Lambda ^4}}}{{{\bf B}_0^2}}} \right)^{2\delta }}, \  \  \
\chi  = -\, 4\, \lambda\, \delta \left( {1 - 2\delta } \right){\left( {\frac{{{\Lambda ^4}}}{{{\bf B}_0^2}}} \right)^{2\delta }}. \ \ \
\label{INVE40}
\end{equation}
Recalling again that we have keep only linear terms in ${\bf E_p}$, ${\bf B_p}$. In this manner, the vacuum electromagnetic properties are clearly characterized by the following expressions for the vacuum permittivity and the vacuum permeability: 
\begin{equation}
{\varepsilon _{ij}} = \xi \,{\delta _{ij}} , \label{INVE45a}
\end{equation}
and
\begin{equation}
{\left( {{\mu ^{ - 1}}} \right)_{ij}} = \xi \left( {{\delta _{ij}} + \frac{\chi }{\xi }\frac{{{B_{0i}}{B_{0j}}}}{{{\bf B}_0^2}}} \right). \label{INVE45b}
\end{equation}

We shall now make a plane wave decomposition for the fields ${\bf E_p}$ and ${\bf B_p}$
\begin{equation}
{{\bf E_p}}\left( {{\bf x}
,t} \right) = {\bf E}\,
{e^{ - i\left( {wt - {\bf k} \cdot {\bf x}} \right)}}, \ \ \
{{\bf B_p}}\left( {{\bf x},t} \right) = {\bf B}\,{e^{ - i\left( {wt - {\bf k} \cdot {\bf x}} \right)}}. \label{INVE50}
\end{equation}
Restricting our considerations to an external magnetic field in the direction $z$, ${\bf B_0} = B_0\, {\bf e}_3$, and the light wave moves along the $x$ axis, we can rewrite the corresponding Maxwell equations in the form
\begin{equation}
\left( {\frac{{{k^2}}}{{{w^2}}} - {\varepsilon _{22}}{\mu _{33}}} \right){E_2} = 0, \label{INVE55a}
\end{equation}  
and
\begin{equation}
\left( {\frac{{{k^2}}}{{{w^2}}} - {\varepsilon _{33}}{\mu _{22}}} \right){E_3} = 0, \label{INVE55b}
\end{equation} 
where we have made use of ${{\bf B}_0} \gg {{\bf B}_p}$.

Next, we also notice that there are two interesting situations from the preceding equations. In fact, 
when ${\bf E}\ \bot \ {\bf B}_0$ (perpendicular polarization), from (\ref{INVE55b}) $E_3=0$, and from
 (\ref{INVE55a}) we get $\frac{{{k^2}}}{{{w^2}}} = {\varepsilon _{22}}\,{\mu _{33}}$. We thus find that the index of refraction is given by 
\begin{equation}
 {n_ \bot } = \sqrt {\frac{{1 + \lambda \left( {1 - 2\delta } \right){{\left( {\frac{{{\Lambda ^4}}}{{{\bf B}_0^2}}} \right)}^{2\delta }}}}{{1 + \lambda \left( {1 - 2\delta } \right)\left( {1 - 4\delta } \right){{\left( {\frac{{{\Lambda ^4}}}{{{\bf B}_0^2}}} \right)}^{2\delta }}}}} . \label{INVE60}
\end{equation}
On the other hand, when ${\bf E}\ || \ {\bf B}_0$ (parallel polarization), from (\ref{INVE55a}) $E_2=0$, and from (\ref{INVE55b}) we get $\frac{{{k^2}}}{{{w^2}}} = {\varepsilon _{33}}\,{\mu _{22}}$. In this case, the index of refraction reduces to
\begin{equation}
{n_\parallel } = 1.  \label{INVE85}
\end{equation}
Accordingly, the preceding electromagnetic vacuum acts like a birefringent medium with two indices of refraction determined by the polarization of the incoming electromagnetic waves. There is, however, a further strand related to the existence of electromagnetic vacuum birefringence. More specifically, we refer to the deflection of a light ray passing through a region where our model under consideration is present. In what follows we will examine this phenomenon.
 
\subsection{Bending of light}

The starting-point of our present discussion is provided by the dispersion equation for an electromagnetic wave propagating in the external magnetic field ${\bf B}_{0}$, for the model given by expression (\ref{INVE05}): 
 \begin{equation}
 {w^2} = {{\bf k}^2}{c^2} - \alpha {c^2}{\left( {{\bf k} \times {\bf B}_{0}} \right)^2},  \label{INVE90}
\end{equation}
 where
 \begin{equation}
\alpha  =   \frac{{\left[ {2\lambda \delta \left( {2\delta  - 1} \right){{\left( {{\raise0.7ex\hbox{${{\Lambda ^4}}$} \!\mathord{\left/
 {\vphantom {{{\Lambda ^4}} 2}}\right.\kern-\nulldelimiterspace}
\!\lower0.7ex\hbox{$2$}}} \right)}^{2\delta }}{{\left( {{\raise0.7ex\hbox{$2$} \!\mathord{\left/
 {\vphantom {2 {{B^2}}}}\right.\kern-\nulldelimiterspace}
\!\lower0.7ex\hbox{${{{\bf B}^2_{0}}}$}}} \right)}^{2\delta  + 1}}} \right]}}{{\left[ {1 - \lambda \left( {2\delta  - 1} \right){{\left( {{\raise0.7ex\hbox{${{\Lambda ^4}}$} \!\mathord{\left/
 {\vphantom {{{\Lambda ^4}} 2}}\right.\kern-\nulldelimiterspace}
\!\lower0.7ex\hbox{$2$}}} \right)}^{2\delta }}} \right]}}.  \label{INVE95}
\end{equation}
The derivation of the preceding dispersion equation may be found in Appendix A.\\

For simplicity, we restrict to $\delta  = 1$ and small $\lambda$. Furthermore, we will consider ${\bf k} \, \bot \, {\bf B}_{0}$. Then, the corresponding dispersion equation read
\begin{equation}
{w^2} = {c^2}\left[ {1 - 2\lambda {{\left( {\frac{{{\Lambda ^4}}}{2}} \right)}^2}{{\left( {\frac{2}{{{{\bf B}_{0}^2}}}} \right)}^3}} \right]{{\bf k}^2}\equiv {c^2}\,\Omega \,{{\bf k}^2}. \label{INVE100}
\end{equation}

Next, in order to obtain the bending angle, we shall work in the ray optics (eikonal) approximation. As is well known, this approximation describes the propagation of light as rays with no reference to the wavelength of the light. We do not attempt here to go in details of this approximation, rather we limit ourselves to writing the ansatz to describe the light as rays, that is, $w = -\frac{{\partial S}}{{\partial t}}$ and ${\bf k} = \nabla S$. Here, $S$ stands for the rapidly varying phase of the electromagnetic wave. Thus, substituting this ansatz according to equation (\ref{INVE100}) we have
\begin{equation}
{\left( {\frac{{\partial S}}{{\partial t}}} \right)^2} - {c^2}\,\Omega {\left( {\nabla S} \right)^2} = 0. \label{INVE105}
\end{equation}

Once again we consider that the external magnetic field is in the direction $z$. In such a case, we may take plane polar coordinates, as shown in Fig. ($1$). Equation (\ref{INVE105}) can therefore be written as follows
\begin{equation}
{\left( {\frac{{\partial S}}{{\partial t}}} \right)^2} - {c^2}\,\Omega \left[ {{{\left( {\frac{{\partial S}}{{\partial r}}} \right)}^2} + \frac{1}{{{r^2}}}{{\left( {\frac{{\partial S}}{{\partial \theta }}} \right)}^2}} \right] = 0. \label{INVE110}
\end{equation}

With the aid of the Hamilton-Jacobi formalism \cite{Denisov}, we write $S = {S_1}\left( r \right) + \alpha\, \theta  - E\,t$, where $\alpha$ is a constant and $E$ is the energy. In this way one encounters 
\begin{equation}
{E^2} - {c^2}\,\Omega \left[ {{{\left( {\frac{{\partial {S_1}}}{{\partial r}}} \right)}^2} + \frac{{{\alpha ^2}}}{{{r^2}}}} \right] = 0. \label{INVE115}
\end{equation}

From the above, we find that $S$ can be brought to the form 
\begin{equation}
S = \int_0^r {d{r^ \prime }} \sqrt {\frac{{{E^2}}}{{{c^2}}}\frac{1}{\Omega } - \frac{{{\alpha ^2}}}{{{r^{ \prime 2}}}}}  + \alpha\, \theta  - E\,t. \label{INVE120}
\end{equation}

Since, $\frac{{dS}}{{d\alpha }} \equiv \beta = constant$, we verify that
\begin{equation}
\beta  = \theta  - {\cal E}\int_0^r {d{r^ \prime }} \frac{1}{{{r^{ \prime 2}}\sqrt {1 + 4\lambda \frac{{{\Lambda ^8}}}{{{{\bf B}^6_{0}}}} - \frac{{{{\cal E}^2}}}{{{r^{ \prime 2}}}}} }}. \label{INVE125}
\end{equation}
Next, differentiating this expression with respect to $r$, and with the substitution $u = \frac{1}{r}$, it follows that
\begin{equation}
\frac{{du}}{{d\theta }} = \frac{1}{{\cal E}}\sqrt {1 + 4\lambda \frac{{{\Lambda ^8}}}{{{{\bf B}^6_{0}}}} - {{\cal E}^2}{u^2}} \label{INVE130}
\end{equation}
where ${\cal E} = \frac{{\alpha c}}{E}$. \\

And, finally, from this expression we obtain
\begin{equation}
sen\left( {\theta  - {\theta _0}} \right) = \frac{{{\cal E}u}}{{\sqrt {1 + 4\lambda \frac{{{\Lambda ^8}}}{{{{\bf B}^6_{0}}}}} }}, \label{INVE135}
\end{equation}
where ${\theta _0}$ is the integration constant. To evaluate ${\theta _0}$ we make use that for $\theta  = \pi$, $u = \frac{1}{{{d_S}}}$. Hence 
\begin{equation}
{\theta _0} = \arcsin \left[ {\frac{{\cal E}}{{{d_S}\sqrt {1 + 4\lambda {{{\Lambda ^8}} \mathord{\left/
 {\vphantom {{{\Lambda ^8}} {{B^6}}}} \right.
 \kern-\nulldelimiterspace} {{{\bf B}^6_{0}}}}} }}} \right]. \label{INVE140}
\end{equation}
Next, we also notice that for ${d_{Ob}} \gg {d_S}$, we have
\begin{equation}
 {d_{Ob}} = arctg\left[ {\frac{{\cal E}}{{{d_S}}}\frac{1}{{\sqrt {1 + 4\lambda {{{\Lambda ^8}} \mathord{\left/
 {\vphantom {{{\Lambda ^8}} {{{\bf B}^6}}}} \right.
 \kern-\nulldelimiterspace} {{{\bf B}^6_{0}}}} - {{{{\cal E}^2}} \mathord{\left/
 {\vphantom {{{{\cal E}^2}} {d_S^2}}} \right.
 \kern-\nulldelimiterspace} {d_S^2}}} }}} \right]. \label{INVE145}
\end{equation}

By defining the bending angle as, $\delta \theta  = {\theta _S} - {\theta _{Ob}}$ (Fig.$1$), we finally obtain the expression
\begin{equation}
\delta \theta  = \arcsin \left[ {\frac{{{\cal E}}}{{{d_S}\sqrt {1 + 4\lambda {{{\Lambda ^8}} \mathord{\left/
 {\vphantom {{{\Lambda ^8}} {{{\bf B}^6}}}} \right.
 \kern-\nulldelimiterspace} {{{\bf B}^6_{0}}}}} }}} \right] - arctg\left[ {\frac{{\cal E}}{{{d_S}\sqrt {1 + 4\lambda {{{\Lambda ^8}} \mathord{\left/
 {\vphantom {{{\Lambda ^8}} {{{\bf B}^6} - {{{{\cal E}^2}} \mathord{\left/
 {\vphantom {{{{\cal E}^2}} {d_S^2}}} \right.
 \kern-\nulldelimiterspace} {d_S^2}}}}} \right.
 \kern-\nulldelimiterspace} {{{\bf B}^6_{0}} - {{{{\cal E}^2}} \mathord{\left/
 {\vphantom {{{E^2}} {d_S^2}}} \right.
 \kern-\nulldelimiterspace} {d_S^2}}}}} }}} \right].  \label{INVE150}
 \end{equation}
 
 An immediate consequence of this is that for $\lambda \to 0$ and  ${d_S} \gg {\cal E}$ one obtains the known Maxwell regime, that is,  $\delta \theta $=0.
 
\begin{figure}[h]
\begin{center}
\includegraphics[scale=1.0]{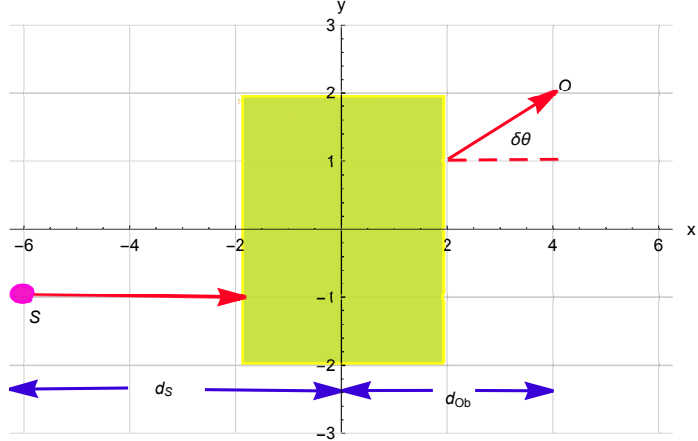}
\end{center}
\caption{\small Bending of light}. \label{horizon}
\end{figure}
 
 \subsection{Interaction energy}
 
 We now proceed to discuss the corrections to the Coulomb potential for the theory under consideration. To do this, we shall compute the expectation value of the energy operator $H$ in the physical state $|\Phi\rangle$ describing the sources, which we will denote by  ${\langle H \rangle}_{\Phi}$. We start then with the Lagrangian density (\ref{INVE05}), that is,
 \begin{equation}
{\cal L} =  - \frac{1}{4}{F_{\mu \nu }}{F^{\mu \nu }} - \frac{\lambda }{4}{F_{\mu \nu }}{F^{\mu \nu }}{\left( {\frac{{2{\Lambda ^4}}}{{{F_{\mu \nu }}{F^{\mu \nu }}}}} \right)^{2\delta }}.  \label{INVE155}
\end{equation}
Making use of $2\delta \ln \left( {\frac{{2{\Lambda ^4}}}{{{F_{\mu \nu }}{F^{\mu \nu }}}}} \right) \ll 1   $, the foregoing Lagrangian density can be brought to the form 
\begin{equation}
{\cal L} =  - \frac{1}{4}\left( {1 + \lambda } \right){F_{\mu \nu }}{F^{\mu \nu }} + \frac{{\lambda \delta }}{2}{F_{\mu \nu }}{F^{\mu \nu }}\ln \left( {\frac{{{F_{\mu \nu }}{F^{\mu \nu }}}}{{2{\Lambda ^4}}}} \right). \label{INVE165}
\end{equation}
To convert the logarithmic term to a more manageable form, we introduce an auxiliary field, $\xi$, \cite{Nonlinear2}. This allows us to write the preceding Lagrangian density as 
\begin{equation}
{\cal L} =  - \frac{1}{4}{A_1}{F_{\mu \nu }}{F^{\mu \nu }} + {A_2}\,{\left( {{F_{\mu \nu }}{F^{\mu \nu }}} \right)^2}, \label{INVE170}
\end{equation}
where ${A_1} = 1 + \lambda  + 2\lambda \delta \left( {1 + \ln \xi } \right)$ and ${A_2} =   \frac{{\lambda \delta \xi }}{{4{\Lambda ^4}}}$. A similar procedure can be used to manipulate the quadratic term in (\ref{INVE170}). Thus, by introducing a second auxiliary field, $\sigma$, we may now write 
\begin{equation}
{\cal L} =  - \frac{1}{4}\sigma {F_{\mu \nu }}{F^{\mu \nu }} - \frac{{{{\left( {\sigma  - {A_1}} \right)}^2}}}{{64{A_2}}}. \label{INVE175}
\end{equation}

Once this is done, canonical quantization is carried out using Dirac's procedure. The canonical momenta are ${\Pi ^\mu } =  - \sigma {F^{0\mu }}$. In this manner we have three primary constraints $\Pi ^0  = 0$, ${{\cal P}_\sigma } \equiv \frac{{\partial L}}{{\partial \dot \sigma }}= 0$ and ${{\cal P}_\xi } \equiv \frac{{\partial L}}{{\partial \dot \xi }}= 0$. The corresponding canonical Hamiltonian is thus
\begin{equation}
{H_C} = \int {{d^3}x} \left\{ {{\Pi _i}{\partial ^i}{A_0} + \frac{1}{{2\sigma }}{{\bf \Pi} ^2} + \frac{\sigma }{2}{{\bf B}^2} + \frac{1}{{64{A_2}}}\left( {\sigma  - A_1} \right)^{2}} \right\}. \label{INVE180}
\end{equation}

Requiring the primary constraint, $\Pi^{0}$, to be stationary, leads to the secondary constraint $\Gamma _1  = \partial _i \Pi ^i  = 0$. It is easily verified that the preservation of $\Gamma_{1}$ for all times does not give rise to any more constraints. In the same way, for the constraint ${{\cal P}_\sigma }$, we get the auxiliary field $\sigma$ as  
\begin{equation}
\sigma  = {\left[ {1 - \lambda  - 2\lambda \delta \left( {1 + \ln \xi } \right) + 12\frac{{\delta \lambda \xi }}{{{\Lambda ^4}}}{{\bf \Pi} ^2}} \right]^{ - 1}}. \label{INVE185}
\end{equation}

Hence we obtain 
\begin{equation}
{H_C} = \int {{d^3}x} \left\{ {{\Pi ^i}{\partial _i}{A_0} + \frac{1}{2}\left[ {1 - \lambda  - 2\lambda \delta \left( {1 + \ln \xi } \right) + 12\frac{{\delta \lambda \xi }}{{{\Lambda ^4}}}{{\bf \Pi} ^2}} \right]{{\bf \Pi} ^2}} \right\}. \label{INVE190}
\end{equation}
It is worthwhile mentioning that to get this last expression we have
ignored the magnetic field in equation (\ref{INVE190}), because it add nothing to the static potential calculation.

Similarly, requiring the constraint, ${{\cal P}_\xi }$, to be stationary in time, we find the auxiliary field $\xi  = \frac{{{\Lambda ^4}}}{{6\,{{\bf \Pi} ^2}}}$. We thus find
\begin{equation}
{H_C} = \int {{d^3}x} \left\{ {{\Pi ^i}{\partial _i}{A_0} + \frac{1}{2}\left[ {1 - \lambda \left( {1 + 4\delta } \right)} \right]{{\bf \Pi} ^2} + \frac{{24\lambda \delta }}{{{\Lambda ^4}}}{{\bf \Pi} ^4}} \right\}. \label{INVE195}
\end{equation}

Next, the extended Hamiltonian  that generates the time evolution of the dynamical variables then reads $H = H_C  + \int {d^3 x} \left( {u_0(x) \Pi_0(x)  + u_1(x) \Gamma _1(x) } \right)$, where $u_o(x)$ and $u_1(x)$ are arbitrary Lagrange multipliers to implement all the first-class constraints. Since $\Pi^0=0$ always and ${\dot {A_0}}\left( x \right) = \left[ {{A_0}\left( x \right),H} \right] = {u_0}\left( x \right)$, which is completely arbitrary, we eliminate $A^0$ and $\Pi^0$ because they add nothing to the description of the system.

Then, the Hamiltonian takes the form 
\begin{equation}
H = \int {{d^3}x} \left\{ {w\left( x \right){\partial _i}{\Pi ^i} + \frac{1}{2}\left[ {1 - \lambda \left( {1 + 4\delta } \right)} \right]{{\bf \Pi} ^2} + \frac{{24\lambda \delta }}{{{\Lambda ^4}}}{{\bf \Pi} ^4}} \right\}, \label{INVE200}
\end{equation}
in which we have used $w(x) = u_1 (x) - A_0 (x)$.
 
It may be noticed here that the quantization of the theory requires the removal of non-physical variables, which is done by imposing a gauge condition such that the full set of constraints become second class. We consequently choose the gauge fixing condition as \cite{Gaete:1997eg}:
 \begin{equation}
\Gamma _2 \left( x \right) \equiv \int\limits_{C_{\zeta x} } {dz^\nu }
A_\nu\left( z \right) \equiv \int\limits_0^1 {d\lambda x^i } A_i \left( {
\lambda x } \right) = 0,  \label{INVE-205}
\end{equation}
where $\lambda$ $(0\leq \lambda\leq1)$ is the parameter describing the
space-like straight path $x^i = \zeta ^i + \lambda \left( {x - \zeta}
\right)^i $ , and $\zeta $ is a fixed point (reference point). We also notice
that there is no essential loss of generality if we restrict our considerations to $\zeta^i=0 
$. Thus, we may now write the only non-vanishing equal-time Dirac bracket, that is,
\begin{equation}
\left\{ {A_i \left( {\bf x} \right),\Pi ^j \left( {\bf y} \right)} \right\}^ * = \delta{_i^j} \,\delta ^{\left( 3 \right)} \left( {{\bf x} - {\bf y}} \right) 
- \partial _i^x 
\int\limits_0^1 {d\lambda \,x^j } \delta ^{\left( 3 \right)} \left( {\lambda
{\bf x}- {\bf y}} \right).  \label{INVE210}
\end{equation}

We are now in a position to obtain the interaction energy by computing the expectation value of the Hamiltonian in the physical state $\left| \Phi  \right\rangle$. We first observe that the physical state, $\left| \Phi  \right\rangle$, is expressed as 
\begin{eqnarray}
\left| \Phi  \right\rangle  \equiv \left| {\bar \Psi \left( {\bf y} \right)\Psi \left( {{{\bf y}^ {\prime} }} \right)} \right\rangle   
= \bar \Psi \left( {\bf y} \right)\exp \left( {iq\int_{{{\bf y}^ {\prime} }}^{\bf y} {d{z^i}{A_i}\left( z \right)} } \right)\Psi \left( {{{\bf y}^ {\prime} }} \right)\left| 0 \right\rangle,   \label{INVE-215}
\end{eqnarray}
where the line integral is along a spacelike path on a fixed time slice, $q$ is the fermionic charge and $\left| 0 \right\rangle$ is the physical vacuum state.

From our above discussion, it follows that
\begin{eqnarray}
{\Pi _i}\left( {\bf x} \right)\left| {\bar \Psi \left( {\bf y} \right)\Psi \left( {{{\bf y}^ {\prime} }} \right)} \right\rangle  &=& \bar \Psi \left( {\bf y} \right)\Psi \left( {{{\bf y}^ {\prime} }} \right){\Pi _i}\left( {\bf x} \right)\left| 0 \right\rangle  
+ q\int_{\bf y}^{{{\bf y}^ {\prime} }} {d{z_i}{\delta ^{\left( 3 \right)}}\left( {{\bf z} - {\bf x}} \right)\left| \Phi  \right\rangle }. 
\label{INVE220}
\end{eqnarray}
In this manner, we obtain the following expectation value of the energy operator  
\begin{equation}
{\left\langle H \right\rangle _\Phi } = {\left\langle H \right\rangle _0} + \left\langle H \right\rangle _\Phi ^{\left( 1 \right)}, 
\label{INVE225}
\end{equation}
where ${\left\langle H \right\rangle _0} = \left\langle 0 \right|H\left| 0 \right\rangle$, whereas the $\left\langle H \right\rangle _0^{\left( 1 \right)}$ term is given by 
\begin{equation}
\left\langle H \right\rangle _\Phi ^{\left( 1 \right)} =   \left\langle \Phi  \right|\int {{d^3}x} \left\{ {\frac{1}{2}\left[ {1 - \lambda \left( {1 + 4\delta } \right)} \right]{{\bf \Pi} ^2} + \frac{{24\lambda \delta }}{{{\Lambda ^4}}}{{\bf \Pi} ^4}} \right\}\left| \Phi  \right\rangle. \label{INVE230}
\end{equation}

Now making use of equation (\ref{INVE220}) and following our earlier procedure \cite{Nonlinear,Logarithmic,Nonlinear2,Nonlinear3}, we find that the potential for two opposite charges, located at ${\bf y}$ and ${\bf y}^{\prime}$, takes the form 
\begin{equation}
V =  - \frac{{{q^2}}}{{4\pi }}\frac{1}{L} + \frac{{3{q^4}}}{{80{\pi ^2}}}\frac{{\lambda \delta }}{{{\Lambda ^4}}}\frac{1}{{{L^5}}}, \label{INVE235}
\end{equation}
where $L = |{\bf y} - {{\bf y}^ \prime }|$. Interestingly, the above static potential profile is analogous to that encountered for generalized Born-Infeld electrodynamics in an external background magnetic field \cite{Accioly}. In this way we have provided a new connection between these effective models.

\section{Related model}

As already stated, our next undertaking is to use the ideas of the previous Section in order to examine a related model of inverse electrodynamics. For this purpose, the authors of ref. \cite{Cembranos} consider the four-dimensional space-time Lagrangian density:
\begin{equation}
{\cal L} =  - \frac{1}{4}{F_{\mu \nu }}{F^{\mu \nu }}\left\{ {1 - \lambda {{\left( {\frac{{{F_{\mu \nu }}{{\tilde F}^{\mu \nu }}}}{{{F_{\mu \nu }}{F^{\mu \nu }}}}} \right)}^2}} \right\}, \label{INVE240}
\end{equation} 
By using, ${\cal F} = \frac{1}{4}{F_{\mu \nu }}{F^{\mu \nu }} = \frac{1}{2}\left( {{{\bf B}^2} - {{\bf E}^2}} \right)$ and ${\cal G} = \frac{1}{4}{F_{\mu \nu }}{\tilde F^{\mu \nu }} =  - {\bf E} \cdot {\bf B}$, the previous Lagrangian density can be written alternatively in the form
\begin{equation}
{\cal L} =  - {\cal F} + \lambda {{\cal F}^{ - 1}}{{\cal G}^2}.\label{INVE245}
\end{equation}

Having made these observations, we can write immediately the field equations
\begin{equation}
\nabla  \cdot {\bf D} = 0, \  \  \
\frac{{\partial {\bf D}}}{{\partial t}} - \nabla  \times {\bf H} = 0, \label{INVE250a}
\end{equation}
\begin{equation}
\nabla  \cdot {\bf B} = 0, \  \  \
\frac{{\partial {\bf B}}}{{\partial t}} + \nabla  \times {\bf E} = 0, \label{INVE250b}
\end{equation}
where the ${\bf D}$ and ${\bf H}$ fields are given by
\begin{equation}
{D_i} = \left[ {\left( {1 + 4\lambda \frac{{{{\left( {{\bf E} \cdot {\bf B}} \right)}^2}}}{{{{\left( {{{\bf B}^2} - {{\bf E}^2}} \right)}^2}}}} \right){E_i} + 4\lambda \frac{{{\bf E} \cdot {\bf B}}}{{\left( {{{\bf B}^2} - {{\bf E}^2}} \right)}}{B_i}} \right], \label{INVE255}
\end{equation}
and
\begin{equation}
{H_i} = \left[ {\left( {1 + 4\lambda \frac{{{{\left( {{\bf E} \cdot {\bf B}} \right)}^2}}}{{{{\left( {{{\bf B}^2} - {{\bf E}^2}} \right)}^2}}}} \right){B_i} - 4\lambda \frac{{{\bf E} \cdot {\bf B}}}{{\left( {{{\bf B}^2} - {{\bf E}^2}} \right)}}{E_i}} \right].  \label{INVE260}
\end{equation}

Following our earlier procedure, we first observe that for a point-like charge, e, at the origin, the 
electrostatic field is given by
\begin{equation}
\left|{\bf E}\right| =\frac{e}{{4\pi {r^2}}}. \label{INVE265}
\end{equation}
Curiously, despite the second term on the right side of (\ref{INVE240}) or (\ref{INVE245}), the previous result indicates that we are in the Maxwell regime. We will elaborate on this point below.

Next, in the same way as was done in the previous Section, the two refractive indices reduce to  
\begin{equation}
{n_\parallel } = \sqrt {1 + 4\lambda }, \label{INVE270}
\end{equation}
and
\begin{equation}
{n_ \bot } = 1. \label{INVE275}
\end{equation}
Here it is important to realize that the corresponding refractive indices are constant and do not depend on the polarization of the external magnetic field. This shows that the vacuum electromagnetic of the present model does not act like a birefringent medium. In fact, this is corroborated by the dispersion equation for the present model, that is, $w^{2} = c^{2}\,{\bf k}^{2}$. The present model therefore describes the Maxwell regime.\\

Before concluding this Section, we discuss the calculation of the interaction energy between static point-like sources for the model under consideration. We shall begin by splitting ${F_{\mu \nu }}$ in the sum of a classical background $\left\langle {{F_{\mu \nu }}} \right\rangle$, and a small fluctuation, $f_{\mu \nu}$. The corresponding Lagrangian density up to quadratic terms in the fluctuations, is given by 
\begin{equation}
{\cal L} =  - \frac{1}{4}{f_{\mu \nu }}{f^{\mu \nu }} - \frac{1}{8}{v^2} + \frac{\lambda }{4}\frac{{\left[ {2{f_{\rho \lambda }}{{\tilde f}^{\rho \lambda }}\left\langle {{F_{\delta \gamma }}} \right\rangle \left\langle {{{\tilde F}^{\delta \gamma }}} \right\rangle  + {{\left( {\left\langle {{F_{\rho \lambda }}} \right\rangle \left\langle {{{\tilde F}^{\rho \lambda }}} \right\rangle } \right)}^2}} \right]}}{{\left( {{f_{\mu \nu }}{f^{\mu \nu }} + \frac{{{v^2}}}{2}} \right)}}, \label{INVE280}
\end{equation}
where ${\varepsilon ^{\mu \nu \alpha \beta }}\left\langle {{F_{\alpha \beta }}} \right\rangle  \equiv {v^{\mu \nu }}$ and $\left\langle {{F_{\mu \nu }}} \right\rangle \left\langle {{F^{\mu \nu }}} \right\rangle  = \frac{{{v^2}}}{2}$.

We first observe that in the ${v^{0i}} \ne 0$ and ${v^{ij}} = 0$ case (referred to as the magnetic one), we find that the density Lagrangian reduces to the Maxwell one
\begin{equation}
{\cal L} =  - \frac{1}{4}{f_{\mu \nu }}{f^{\mu \nu }}, \label{INVE285}
\end{equation}
to get the last line we have made use of $\left\langle {{F^{\delta \gamma }}} \right\rangle \left\langle {{{\tilde F}_{\delta \gamma }}} \right\rangle  = 0$, and we have ignored the constant $\frac{1}{8}{v^2}$.

Second, in the ${v^{0i}}=0$ and ${v^{ij}} \ne 0$ case, we also obtain $\left\langle {{F^{\delta \gamma }}} \right\rangle \left\langle {{{\tilde F}_{\delta \gamma }}} \right\rangle  = 0$,
leading to the Maxwell theory.

In summary then, in both cases, the static potential profile reduces to the Coulombic one \cite{Gaete98}.

\section{Final remarks}

In summary, in this work we have considered a new nonlinear electrodynamics. It was shown that in this new electrodynamics the phenomenon of birefringence takes place in the presence of an external magnetic field. Subsequently, we have obtained an expression for the bending of light. Afterwards, we have calculated the interaction energy. As in previous works  \cite{Nonlinear,Logarithmic,Nonlinear2,Nonlinear3}, we have exploited a correct identification of field degrees of freedom with observable quantities. Interestingly, it was shown that the static potential profile contains a long-range (${1 \mathord{\left/{\vphantom {1 {{r^5}}}} \right.\kern-\nulldelimiterspace} {{r^5}}}$-type) correction, to the Coulomb potential. In addition, we have compared our previous results with a related model of inverse electrodynamics. It remains to be worked out how to connect our results with ongoing experiments related to light-by-light nonlinearity effects. We shall be reporting on that in a forthcoming work.

\section{Acknowledgement}

One of us (P. G.) was partially supported by Fondecyt (Chile) grant 1180178 and by ANID PIA / APOYO AFB180002.

\section{Appendix A: Dispersion relation}

In this Appendix, we address the problem of deriving the dispersion relation given by equation (\ref{INVE100}).
To do that we consider a generic Lagrangian density:
\begin{equation}
{\cal L} = {\cal L}({\cal F},{\cal G}), \label{App05}
\end{equation}
where ${\cal F} =  - \frac{1}{4}{F^{\mu \nu }}{F_{\mu \nu }} = \frac{1}{2}\left( {({{\bf E}^2} - {{\bf B}^2}} \right)$ and 
${\cal G} =  - \frac{1}{4}{F^{\mu \nu }}{\tilde F_{\mu \nu }} = {\bf E} \cdot {\bf B}$.

Next, after splitting ${F^{\mu \nu }}$ in the sum of a classical background, $F_B^{\mu \nu }$, and a small fluctuation, ${f^{\mu \nu }}$, the corresponding linearized field equations read 
\begin{equation}
{\partial _\mu }\left( {{C_1}{f^{\mu \nu }} + {C_2}{{\tilde f}^{\mu \nu }}} \right) - \frac{1}{2}{\partial _\mu }\left( {k_B^{\mu \nu \kappa \lambda }{f_{\kappa \lambda }} + t_B^{\mu \nu \kappa \lambda }{{\tilde f}_{\kappa \lambda }}} \right) - \frac{1}{4}{\partial _\mu }\left( {{\varepsilon ^{\mu \nu \kappa \lambda }}{t_{B\kappa \lambda \rho \sigma }}{f^{\rho \sigma }}} \right)
 =  - {\partial _\mu }\left( {{C_1}F_B^{\mu \nu } + {C_2}{{\tilde F}^{\mu \nu }}} \right) + {j^\nu }, \label{App10}
\end{equation}
where $k_B^{\mu \nu \kappa \lambda } = {D_1}F_B^{\mu \nu }F_B^{\kappa \lambda } + {D_2}\tilde F_B^{\mu \nu }\tilde F_B^{\kappa \lambda }$ and $t_B^{\mu \nu \kappa \lambda } = {D_3}F_B^{\mu \nu }F_B^{\kappa \lambda }$. Whereas ${C_1} = {\left. {\frac{{\partial {\cal L}}}{{\partial {\cal F}}}} \right|_B}$, ${C_2} = {\left. {\frac{{\partial L}}{{\partial {\cal G}}}} \right|_B}$, ${D_1} = {\left. {\frac{{{\partial ^2}{\cal L}}}{{\partial {{\cal F}^2}}}} \right|_B}$, ${D_2} = {\left. {\frac{{{\partial ^2}{\cal L}}}{{\partial {{\cal G}^2}}}} \right|_B}$ and ${D_3} = {\left. {\frac{{{\partial ^2}{\cal L}}}{{\partial {\cal F}\partial {\cal G}}}} \right|_B}$.

However, in what follows we will compute the dispersion relation in the case ${\cal L}={\cal L}({\cal F})$ and $j^{\nu}=0$, in the presence of a constant background with both electric and magnetic fields $({\bf E},{\bf B})$. 
\begin{eqnarray}
\nabla  \cdot {\bf e} + \frac{{{D_1}}}{{{C_1}}}{\bf E} \cdot \nabla \left( {\frac{1}{{{c^2}}}{\bf E} \cdot {\bf e} - {\bf B} \cdot {\bf b}} \right) = 0,\,\,\,\,\,\,\frac{{\partial {\bf b}}}{{\partial t}} =  - \nabla  \times {\bf e}, \nonumber\\
\nabla  \cdot {\bf b} = 0,\,\,\,\,\, \nabla  \times {\bf b} + \frac{{{D_1}}}{{{C_1}}}{\bf B} \times \nabla \left( {{\bf B} \cdot {\bf b} - \frac{1}{{{c^2}}}{\bf E} \cdot {\bf e}} \right) = \frac{1}{{{c^2}}}\frac{{\partial {\bf e}}}{{\partial t}} + \frac{1}{{{c^2}}}{\bf E}\frac{\partial }{{\partial t}}\left( {\frac{1}{{{c^2}}}{\bf E} \cdot {\bf e} - {\bf B} \cdot {\bf b}} \right).
\label{App15}
\end{eqnarray}
Throughout, ${\bf e}$ and ${\bf b}$ are the electric and magnetic fields arising from the fluctuation ${f^{\mu \nu }}$.

By considering the plane waves
\begin{equation}
{\bf e} = {{\bf e}_0}\,{e^{i\left( {{\bf k} \cdot {\bf x} - wt} \right)}}, \,\,\,\,\,{\bf b} = {{\bf b}_0}\,{e^{i\left( {{\bf k} \cdot {\bf x} - wt} \right)}}, \label{App20}
\end{equation}
from the equations (\ref{App15}) it follows that 
\begin{equation}
{M_{ij}}{e_{0j}} = 0, \label{App25}
\end{equation}
where 
\begin{equation}
 {M_{ij}} = \left( {\frac{{{w^2}}}{{{c^2}}} - {{\bf k}^2}} \right){\delta _{ij}} + {k_i}{k_j} + \frac{{{D_1}}}{{{C_1}}}{\Omega _i}{\Omega _j}, \label{App30}
\end{equation}
in which we have used ${\bf \Omega}  \equiv \frac{w}{{{c^2}}}{\bf E} + {\bf k} \times {\bf B}$. It should be further recalled that such a system of linear homogeneous equations (\ref{App25}) whose solution is not zero only if the determinant of its coefficients vanishes, that is, $detM=det[M_{ij}]=0$.
Rewriting equation (\ref{App30}) in the form ${M_{ij}} = a{\delta _{ij}} + b{u_i}{u_j} + d{v_i}{v_j}$, we find for the $detM$ the expression
\begin{equation}
\det M = a\left[ {\left( {a + b{{\bf u}^2}} \right)\left( {a + d{{\bf v}^2}} \right) - bd{{\left( {{\bf u} \cdot {\bf {\bf v}}} \right)}^2}} \right], \label{App35}
\end{equation}
where $a = \frac{{{w^2}}}{{{c^2}}} - {{\bf k}^2}$, $b=1$, $d=\frac{{{D_1}}}{{{C_1}}}$, ${\bf u}={\bf k}$ and ${\bf v}={\bf \Omega}$. From this expression we readily deduce that 
\begin{equation}
{w^2}\left( {\frac{{{w^2}}}{{{c^2}}} - {{\bf k}^2}} \right)\left[ {\alpha \frac{{{w^2}}}{{{c^2}}} + \beta \frac{w}{c} + \gamma } \right] = 0, \label{App40}
\end{equation}
where
\begin{equation}
\alpha  = \left( {1 + \frac{{{D_1}}}{{{C_1}}}\frac{{{{\bf E}^2}}}{{{c^2}}}} \right), \,\,\,\,
\beta  = 2\frac{{{D_1}}}{{{C_1}}}\frac{{\bf E}}{c} \cdot \left( {{\bf k} \times {\bf B}} \right),\,\,\,\, \gamma  = \left( { - {{\bf k}^2} + \frac{{{D_1}}}{{{C_1}}}{{\left( {{\bf k} \times {\bf B}} \right)}^2} - \frac{{{D_1}}}{{{C_1}}}{{\left( {{\bf k} \cdot \frac{{\bf E}}{c}} \right)}^2}} \right). \label{App45}
\end{equation}
Evidently, $w^{2}=0$ is meaningless and $w^{2}={\bf k}^{2}c^{2}$ describes the Maxwell regime.\\

For the model under consideration, we have ${\bf E}=0$, therefore $\alpha=1$, $\beta=0$ and $\gamma= - {{\bf k}^2} + \frac{{{D_1}}}{{{C_1}}}{{\left( {{\bf k} \times {\bf B}} \right)}^2}$. In addition, we have 
\begin{equation}
\frac{{{D_1}}}{{{C_1}}} =   \frac{{\left[ {2\delta \left( {2\delta  - 1} \right)\lambda {{\left( {\frac{{{\Lambda ^4}}}{2}} \right)}^{2\delta }}{{\left( {\frac{2}{{{{\bf B}^2}}}} \right)}^{2\delta  + 1}}} \right]}}{{\left[ {1 - \lambda \left( {2\delta  - 1} \right){{\left( {\frac{{{\Lambda ^4}}}{{{{\bf B}^2}}}} \right)}^{2\delta }}} \right]}}.  \label{App50}
\end{equation}
In the case $\delta=1$ and small $\lambda$ we obtain
\begin{equation}
\frac{{{D_1}}}{{{C_1}}} \simeq   2\lambda {\left( {\frac{{{\Lambda ^4}}}{2}} \right)^2}{\left( {\frac{2}{{{{\bf B}^2}}}} \right)^3}. \label{App55}
\end{equation}
Using that ${\bf k} \bot {\bf B}$, we accordingly express the dispersion relation in the form 
\begin{equation}
{w^2} = {c^2}\left[ {1 - 2\lambda {{\left( {\frac{{{\Lambda ^4}}}{2}} \right)}^2}{{\left( {\frac{2}{{{{\bf B}^2}}}} \right)}^3}} \right]{{\bf k}^2}, \label{App55}
\end{equation}
which coincides with equation $(\ref{INVE100})$.

\end{document}